\documentclass[runningheads]{llncs}
\usepackage[T1]{fontenc}
\usepackage{graphicx}
\usepackage{algorithmic}
\usepackage{graphicx}
\usepackage{textcomp}
\usepackage{xcolor}
\usepackage{comment}

\usepackage{caption}
\usepackage{subcaption}
\usepackage{multirow}
\usepackage{adjustbox}
\usepackage{balance}
\usepackage{soul}
\usepackage{hyperref}
\usepackage{url}

\usepackage{ctable}

\usepackage{placeins} 

\usepackage{changepage} 

\newcommand{\squishlist}{
 \begin{list}{$\bullet$}
  { \setlength{\itemsep}{0pt}
     \setlength{\parsep}{2pt}
     \setlength{\topsep}{2pt}
     \setlength{\partopsep}{0pt}
     \setlength{\leftmargin}{1em}
     \setlength{\labelwidth}{0.5em}
     \setlength{\labelsep}{0.3em} } }

\newcommand{\squishend}{
  \end{list}  }

\definecolor[named]{cool}{RGB}{59, 77, 191}
\definecolor[named]{warm}{RGB}{180, 4, 39}

\definecolor{notext}{rgb}{0.64, 0.64, 0.82}

\definecolor{cadmiumorange}{rgb}{0.93, 0.53, 0.18}
\definecolor{darkorchid}{rgb}{0.6, 0.2, 0.8}
\definecolor{brilliantrose}{rgb}{1.0, 0.33, 0.64}

\begin{document}
\title{Recommendations by Concise User Profiles from Review Text}

\author{Ghazaleh H. Torbati\inst{1}
\and
Anna Tigunova\inst{2}
\and
Gerhard Weikum\inst{1}
\and
Andrew Yates\inst{3,4}
}

\institute{Max Planck Institute for Informatics, Saarbrücken, Germany \\
\email{\{ghazaleh,weikum\}@mpi-inf.mpg.de} \and
Amazon, Germany \\
\email{tigunova@amazon.com} 
\and University of Amsterdam, Amsterdam, Netherlands
\and HLTCOE, Johns Hopkins University, Baltimore, Maryland, USA \\
\email{andrew.yates@jhu.edu}
}

\maketitle              
\begin{abstract}
Recommender systems perform well for popular items and users with ample interactions (likes, ratings etc.).
This work addresses the difficult and underexplored case of users who have very sparse interactions but post informative review texts.
This setting naturally calls for encoding user-specific text with large language models (LLM).
However, feeding the full text of all reviews through an LLM has a weak signal-to-noise ratio and incurs high costs of processed tokens.
This paper addresses these two issues.
It presents a light-weight framework, called CUP, which first computes concise user profiles and feeds only these 
into the training of transformer-based recommenders.
For user profiles,
we devise various techniques 
to select the most informative cues from noisy reviews.
Experiments, with book reviews data, show that
fine-tuning a small language model with 
judiciously constructed profiles
achieves the best performance, even in comparison to LLM-generated rankings.
\end{abstract}
\section{Introduction}

\noindent{\bf Motivation:}
\label{sec:intro-motivation}
Recommender-system 
methods fall into two major families or hybrid combinations \cite{DBLP:reference/sp/2022rsh}:
i) {\em interaction-based} recommenders 
that leverage binary signals 
(e.g., membership in personal playlists or libraries) 
or numeric ratings for user-item pairs, and
ii) {\em content-based} recommenders that exploit item features and user-provided content, ranging
from metadata attributes (e.g., item categories) all the way to review texts.

In settings where interaction data is sparse, content-based methods are the only option, and this is the focus of this work.
The most promising approach for this regime is to harness {\em review texts} by users (e.g., \cite{DBLP:conf/www/ChenZLM18,DBLP:conf/kdd/LiuLDCG19,DBLP:journals/corr/abs-2104-13030,DBLP:journals/csur/ZhangYST19}).
In domains where users spend substantial time per item (e.g., books, travel destination), unlike short-attention-span items (e.g., music, video streams), users tend to leave detailed reviews expressing their interests and tastes, even with few interactions.
This paper 
presents a new framework to tackle 
review-based recommendation
with sparse data, long-tail users and items, and rich review texts, especially when computational resources are limited.
We focus on the domain of books as a prime case for
low interaction rates (i.e., users with few items) with high interaction efforts (i.e., value in user reviews), in combination with high diversity of user tastes (both across users and also per user). 
Although this is not in the mainstream 
business, we advocate 
that long-tail users 
should receive better service as well.


\vspace{0.1cm}
\noindent{\bf State-of-the-Art Limitations:}
Recent works integrated
item descriptions and
textual reviews into various kinds of recommender architectures, 
including some 
based on
large language 
models (LLMs) (e.g., \cite{DongFangLauw:WSDM2025,DBLP:journals/corr/abs-2402-11143,DBLP:conf/www/LinCWXQDZTY024,DBLP:conf/acl/RamosRWFL24}).
Our approach differs fundamentally, by making user profiles explicit and transparent, before feeding them into a recommender. This way, lay users can inspect, edit, extend or customize their profiles in a human-friendly manner,
while personalizing the downstream application.

There are established works on cold-start support and long-tail items (e.g., \cite{DBLP:conf/www/CaoWHH20,DBLP:conf/aaai/LiJL00H19,DBLP:conf/kdd/LiuBXG022,DBLP:conf/cikm/LuoMXS23,DBLP:conf/recsys/Raziperchikolaei21}). These are driven by similarity-aware architectures such as graph models, matrix factorization, and neural methods. Their key asset is to infer explicit or implicit properties of long-tail items and the resulting user preferences, by learning from similar items with richer data. This approach does not carry over to long-tail {\em users}, though, 
when most users have sparse data and high diversity in tastes and interests.
In book communities such as Goodreads, we encounter many users with just 
a few tens of items across widely different genres.

\begin{figure}[h]
    \centering
    \begin{minipage}{0.55\textwidth}
        \centering
        \includegraphics[width=\linewidth]{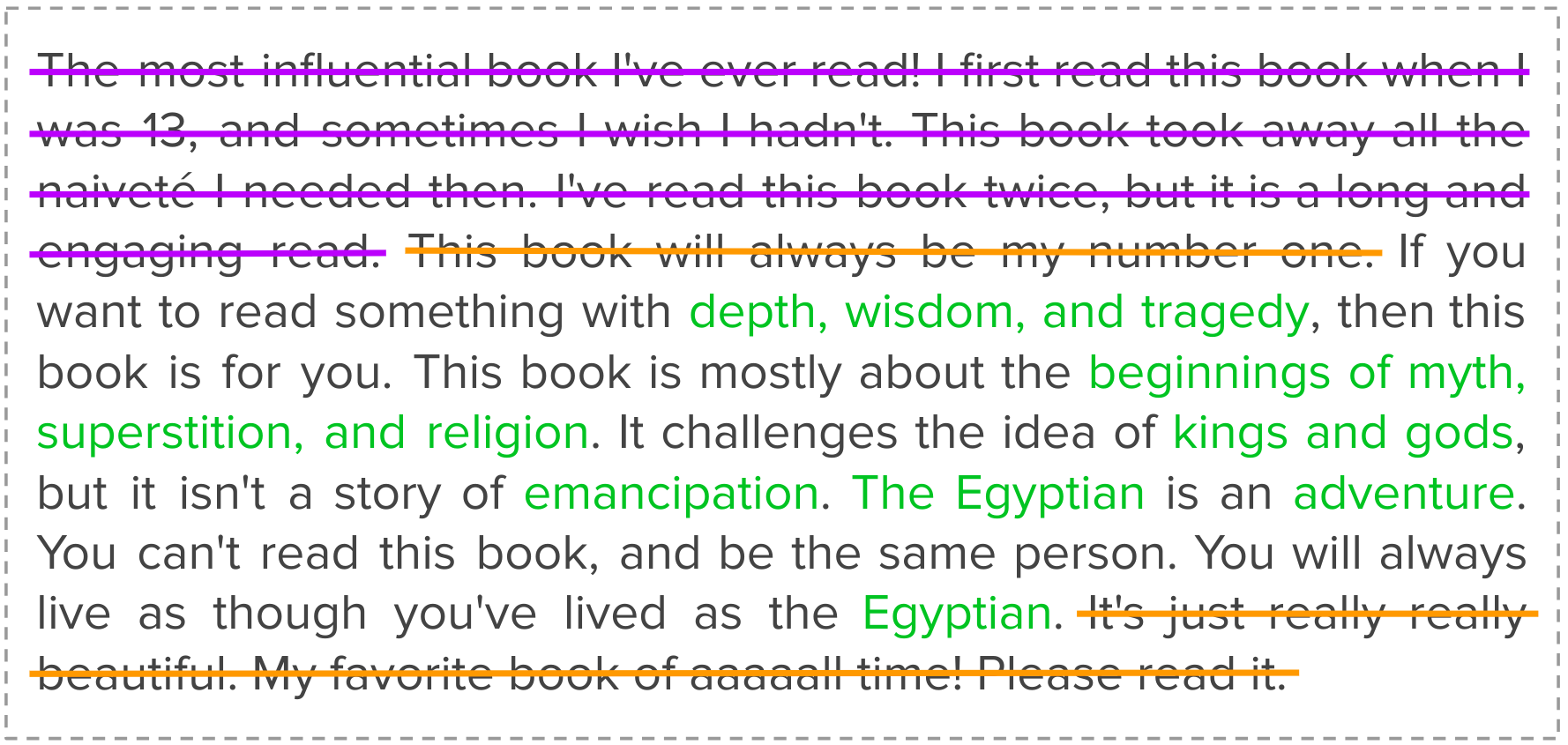}  
        \caption{\small User-written review, with uninformative text crossed over. Personal background is in {\color{purple} purple}, pure sentiment in {\color{cadmiumorange} orange}, most informative cues in {\color{green} green}.}
        \label{fig:reviewexample}
    \end{minipage}%
    \hfill
    \begin{minipage}{0.44\textwidth}
    \centering
    \captionof{table}{Datasets statistics.}
    \label{tab:dataset-stats}
    {\small
    \resizebox{\textwidth}{!}{
        \begin{tabular}{|l|ccccc|} \hline
         & \rotatebox{90}{\# items} &  \rotatebox{90}{\# users per i} & \rotatebox{90}{\# users} & \rotatebox{90}{\# items per u} & \rotatebox{90}{avg review len} \\ \hline
GR              & 1.5M   & 6.54   & 280K  & 36.77     & 178 \\ 
GR-1K-rich      & 45K    & 1.17  & 1K    & 53.32     & 426 \\ 
AM              & 2.3M   & 12.75  & 3M    & 9.37      & 121 \\ 
AM-1K-rich      & 16K   & 1.07   & 1K    & 16.79     & 282 \\ \hline
\end{tabular}
    }
    }
    \end{minipage}
\end{figure}

\vspace{0.1cm}
\noindent{\bf Research Challenges:}
\label{subsec:intro-RQ}
Our approach constructs {\em concise user profiles} from rich but noisy reviews, and feeds these into the training of a two-tower transformer-based recommender system. 
This comes with three main challenges:
{{\em 1) Long-tail Users:}}
Unlike for long-tail items, there is no way for transferring knowledge from dense-interaction {\em users} to users in the long tail. The sparseness and the high diversity of user interests and tastes pose unique challenges.
{{\em 2) Noisy Reviews:}}
User reviews express a mix of aspects:
personal background, 
interesting traits of the reviewed item, and
sentiment expressions.
Figure \ref{fig:reviewexample} depicts an example review with a mix of informative and uninformative signals.
The challenge lies in identifying the 
scarce parts of a review that convey information about the {\em item itself} and {\em why} the user likes (or dislikes) it.
{{\em 3) Low-resource Computation:}}
LLMs can handle large inputs, but the computational and energy cost increases with the number of input tokens. 
The challenge is to get high mileage while keeping the footprint and computation low.

\vspace{0.1cm}
\noindent{\bf Approach:}
We devise a light-weight framework, called {\bf CUP}, for constructing $\underline{\rm C}$oncise
$\underline{\rm U}$ser $\underline{\rm P}$rofiles
and encoding them in a recommender system.
We adopt a two-tower transformer-based architecture that supports end-to-end learning of item and user encodings,
making use of a (small) language model (LM).
The end-to-end learning operates 
on short, judiciously constructed profiles.
Our choice for the  encoder is BERT; alternatives such as T5, GPT or Llama can be easily plugged in.

On top of the transformer, we place feed-forward layers, which provide controllable fine-tuning for the downstream recommendation task.
The 
prediction scores for user-item pairs
yield the per-user ranking of items.
This architecture is relatively simple, but very versatile in supporting different configurations
and being able to incorporate a wide range of user profiling techniques.

Our experiments compare against several baselines, including LLM-based ranking.
We focus on the book domain, using slices of datasets from Amazon and Goodreads, where we select users with long review text (see Table \ref{tab:dataset-stats} for dataset statistics).
Code and data are available at {\small\url{https://personalization.mpi-inf.mpg.de/CUP}}. 

\vspace*{0.1cm}
\noindent{\bf Contributions:}
Salient contributions of this work are:
i) a new framework, called CUP, for transformer-based recommenders that leverage 
concise user profiles from reviews;
ii) judicious techniques for selecting and encoding informative cues from long and noisy reviews;
iii) comprehensive experiments with data-poor but text-rich users with highly diverse preferences.


\vspace{-0.4cm}

\vspace{-0.2cm}
\section{Related Work}
\vspace{-0.2cm}

\noindent{\bf Exploiting User Reviews:}
Incorporating user-provided reviews into recommenders
has been pursued with deep neural networks 
\cite{DBLP:conf/www/ChenZLM18,DBLP:conf/kdd/LiuLDCG19,wu2019neural,DBLP:journals/corr/abs-2104-13030,DBLP:conf/cikm/ZhangACC17,10.1145/3018661.3018665}
and latent-factor models \cite{DBLP:conf/www/HuZY19,DBLP:conf/recsys/PenaOTHDSL20,DBLP:conf/recsys/ShalomUK19}.
Some works augment collaborative filtering (CF) models with user text, to mitigate data sparseness
(e.g., \cite{liu2020hybrid,lu2018coevolutionary,sar2018word}).
\cite{DBLP:conf/www/0012OM21} proposes to learn the 
importance of review meta-data (age, length etc.), but textual content is disregarded.
Other works like \cite{DBLP:conf/sigir/ShuaiZWSHWL22} incorporate the similarity of user reviews and item descriptions into graph-based learning.
Mostly pre-dating the advent of 
large language models, these methods 
have been 
found to have only
limited effects \cite{DBLP:conf/sigir/SachdevaM20}.

Recent work by \cite{DBLP:conf/acl/RamosRWFL24,DBLP:conf/wsdm/TorbatiTW24} takes advantage of large language models to generate short user profiles from 
reviews and item descriptions. Our framework subsumes this approach as a special case.
The brand-new work of \cite{DongFangLauw:WSDM2025} 
includes reviews 
in learning a model 
over a shared latent space. 
In contrast, our framework makes user profiles explicit before feeding them into the recommender, giving the user a chance to inspect and edit this personal data.

\vspace{0.1cm}
\noindent{\bf Exploiting Language Models:}
Pre-trained LMs can be leveraged 
to i) encode item-user signals into transformer-based embeddings, ii) infer recommended items from rich representations of review texts, or iii) implicitly incorporate the latent ``world knowledge'' of the LM. 

A represenative of the first line is P5 \cite{DBLP:conf/recsys/Geng0FGZ22},
which
employs prompt templates for the T5 language model.
We include an 
{enhanced}
variant of the P5 method in our experiments.
The recent work of \cite{DBLP:conf/acl/RamosRWFL24}
generates user profiles by prompting LLMs like Llama, Mistral. 
The profiles are used for rating prediction, not for ranking a larger pool of candidate items. 
The method is designed for short reviews; text-rich book reviews are not considered.

On the second direction, the work of \cite{pugoy-kao-2020-bert,pugoy-kao-2021-unsupervised},
uses BERT to create representations for user and item text, with (short chunks of) single reviews as granularity.
The method then aggregates these per-review vectors by averaging 
\cite{pugoy-kao-2020-bert} 
or k-means clustering \cite{pugoy-kao-2021-unsupervised}.
Our experiments include BENEFICT 
\cite{pugoy-kao-2020-bert} as a baseline.

On the ``world knowledge'' direction, 
early works, using BERT, elicit knowledge about movie, music and book genres \cite{DBLP:conf/recsys/PenhaH20}.  Recent works prompt large language models (LLMs), such as GPT or Llama, to generate item rankings for user-specific recommendations \cite{hou2023large,wang2023zero} or predict user ratings \cite{kang2023llms}, in a zero-shot 
or few-shot 
fashion. 
{Our experiments include \cite{hou2023large} as an LLM-powered baseline.}


\vspace{0.1cm}
\noindent{\bf Supporting the Long Tail:}
Support for long-tail items and users falls under the theme of cold-start and zero-shot recommendations (e.g., \cite{DBLP:conf/www/CaoWHH20,DBLP:conf/aaai/LiJL00H19,DBLP:conf/kdd/LiuBXG022,DBLP:conf/cikm/LuoMXS23,DBLP:conf/recsys/Raziperchikolaei21,DBLP:journals/corr/abs-2108-03357}).
State-of-the-art methods are reasonably successful on new items, by embedding the item features into the same space as warm items, thus learning relatedness between warm and cold items. This assumes that cold items come with tags and descriptions. For the user side, this assumption is not practical:
users would not likely expose a rich profile when they are new to a community or merely occasional contributors.
In this data-poor regime, the only option is to harness textual cues from a small number of reviews.

\vspace{-0.2cm}
\section{Methodology}
\vspace{-0.2cm}
\subsection{System Architecture}
The CUP framework is based on a two-tower architecture for representation learning (one "tower" for users, the other for items, following the prevalent architecture in neural information retrieval with query and document/passage encodings). The two towers are jointly trained, coupled by the
shared loss function.

User profile and item metadata
are fed into BERT followed by a feed-forward network to learn latent representations.
Downstream, the vectors are simply compared by a dot product for scores that indicate whether the user likes an item or not. 
The {\em per-item} text usually comprises
book titles, tags like categories or genre labels,
which can be coarse (e.g., ``thriller'') or fine-grained (e.g., ``Scandinavian crime noir''), and a short description of the book contents.
The {\em per-user} text can comprise the
titles and tags of her training-set books 
and the entirety of her review texts, 
which vary widely in length
{and informativeness}, hence the need for 
{smart text selection}.

\vspace{-0.2cm}
\subsection{Training}
\vspace{-0.2cm}
\label{subsec:training}
The input to CUP consists of an item metadata 
(almost always short, otherwise truncated) and a judiciously selected subset of the user-provided text (which may total to a longer text).
For user $u$, this is a sequence of text tokens $w^u_1 \dots w^u_b$, where $b$ is the token budget by which the input is limited (set to 128 in our experiments). The sequence is fed through the user tower, consisting of BERT encoder and a feed-forward network (FFN), to obtain a user-representation vector $t^u$
(by averaging the per-token vectors).
The FFN has two layers with ReLU activation, 
computing the final user representation as
$~~ t^u = ReLU(t^u W^u_1 + c^u_1)W^u_2 + c^u_2 ~${;}
and analogously for items.
The score for a \textit{user-item pair} 
is calculated by
$~~ s^{ui} = \sigma(<t^u, t^i>) ~$
with  dot product $<,>$ and sigmoid function $\sigma$.

We use the Adam optimizer to minimize the binary cross-entropy loss between predicted labels and the ground truth 
{with sampled negatives}.
During training we update the top-most layer of BERT, which allows end-to-end training of all components. 

\vspace{-0.2cm}
\subsection{Inference}
\vspace{-0.1cm}
\label{subsec:inference}

\noindent{\bf Prediction for Ranking.}
At test time, a prediction is made for user-item pairs.
We encode the item description by running it through the trained network, and we compare it to the already learned user vector, which is based on the user's training-time reviews.
The scores for different test items, computed by the final dot product, yield the ranking of the candidate items.
This is a very efficient computation, following
established practice in neural {IR} 
\cite{DBLP:series/synthesis/2021LinNY}.

\noindent{\bf Search-based Recommendation.}
In a deployed system (as opposed to lab experiments with test samples), a typical usage mode would be
search-based re-ranking: a user 
provides context with a 
keyphrase query or an example of a specific liked item,
which can be thought of as query-by-example. 
The user's expectation
is to see a ranked list of recommended items that are similar to 
her {search intent}  
(as opposed to recommendations from all kinds of categories).
The system achieves this by first computing 
approximate matches 
to the query 
(i.e., similarity-search neighbors), and then
re-ranking a shortlist of say top-100
candidates.
The CUP framework supports this mode, by using a
light-weight BM25 retrieval model. 
To evaluate the model in this mode without ground truth query-user-item triples, we search over all unlabeled items
with the category and textual description of the positive {test} point at hand, and keeping the 
top-100 highest scoring matches.

\vspace{-0.2cm}
\subsection{Coping with Long and Noisy Texts}
\vspace{-0.2cm}
\label{subsec:longtext}

For constructing user profiles from text, the simplest idea would be
to concatenate all available reviews into a long token sequence.
Two problems arise, though. 
User reviews are a noisy mix of descriptive elements (e.g., ``the unusual murder weapon''), sentiment expressions (e.g., ``it was fun to read'') and
personal but irrelevant statements (e.g., ``I read only on weekends'').
Only the first aspect is helpful for content-based profiling (as the sentiment is already captured by user liking the book).
Second, the entirety of user-provided text can be too long to be fully digested by the Transformer. Even when it would fit into the token budget, the computational and energy costs are quadratic in the number of input tokens. 
Therefore, we tightly limit the tokens for each user's text profile to 128, and devise a suite of light-weight techniques for judiciously selecting the most informative pieces.

Our techniques for selecting the most informative parts of user reviews 
into a concise user profile
are as follows:
\squishlist
\item {\bf Weighted Phrases:} selected words
or 3-grams, ordered by descending tf-idf weights, where tf 
is the frequency {of the phrase} in all of the user's reviews,
and idf is pre-computed on Google books n-grams
to capture informativeness. 
\item {\bf Weighted Sentences:} 
selected sentences, ordered by descending idf weights,
where
a sentence's total weight is the sum of the per-word idf weights normalized by sentence length.
\item {\bf Similar Sentences:} selected sentences, ordered by descending similarity scores computed via Sentence-BERT \cite{DBLP:conf/emnlp/ReimersG19} for comparing the {user-review} 
sentences against 
{the description of the corresponding item.}
To ensure that the selected set is not dominated by a single review, the sentences are picked from different reviews in a round-robin manner.
\item {\bf ChatGPT-generated Profiles:}
feeding all reviews of a user, in large chunks, 
into ChatGPT and {instructing} it to characterize the user's {book} interests 
{with a few keyphrases.}
\item {\bf T5-generated Keywords:} 
using a T5 model fine-tuned for keyword generation,
to cast each user's review text into a set of keywords, concatenated to create the profile.
\item {\bf Llama-generated Profiles:} 
instructing Llama \cite{DBLP:journals/corr/abs-2407-21783} to generate profiles given the user reviews in two formats: keywords, and first-person narrative, by giving it a few in-context examples.
\squishend
We 
provide
anecdotal 
examples of selected user profile constructions
in Table \ref{table:profiles}.
%


\vspace{-0.2cm}
\subsection{Coping with Unlabeled Data}
\vspace{-0.2cm}
\label{subsec:methodology-unlabeleddata}
A challenge for training in the data-poor regime is handling the extreme skew between positive samples and unlabeled data points for sparse users. The crux in many recommender applications is that there are extremely few, if any, explicitly negative samples, such as books rated with low marks. This holds also for the datasets in this work. 
Therefore, we 
introduce and 
experiment with two different techniques to construct negative training samples from unlabeled data:

\noindent{\bf Uniform random samples.}
Under the closed world assumption (CWA), aka Selected Completely At Random, negative training points
are sampled uniformly from all unlabeled data.
This is a widely used standard technique. 

\noindent{\bf Weighted pos-neg samples.}
Prior works on PU learning \cite{DBLP:journals/ml/BekkerD20}, with positive and unlabeled data and without explicitly negative samples, is largely based on treating unlabeled points as pairs of samples, one positive and one negative with fractional weights. The weights can be based on (learned estimates of) class priors, but the extreme skew in our data renders these techniques 
ineffective.
Instead, we leverage the fact that 
relatedness measures between item pairs can be derived from
interaction data.
We 
compute {\em item-item relatedness} via 
matrix factorization of the user-item matrix for the entire dataset.
The relatedness of two items is set to the scalar product between their latent vectors, 
re-scaled for normalization
between 0 and 1.
Each originally unlabeled sample is cloned:
one instance positive with weight
proportional to its average relatedness to the
user's explicitly positive points;
the negative clone's weight is set to the complement.

\vspace{-0.2cm}
\section{Experimental Design}
\label{sec:experiments-design}
\vspace{-0.2cm}

\noindent{\bf Rationale:}
As a difficult 
and less explored application area for recommenders, we investigate the case of book recommendations in online communities. 
These come with a long-tailed distribution of user activities, highly diverse user interests, 
and demanding textual cues from reviews and book descriptions.

Unlike in many prior works' experiments, often
on movies, restaurants or mainstream products,
the data in our experiments is much sparser regarding user-item interactions. 
We design the evaluation as a {\em stress-test} experiment, with
focus on text-rich
users.
With the focus on lightweight computation, we limited the input context to 128 tokens, hence the need for smart user profile extraction. Our experiments supports the choice of budget and architecture.  

We further
enforce the {\em difficulty of predictions} when items belong to groups with high relatedness within a group, 
by constraining disjointedness of authors per user in training and evaluation set.
Thus, we rule out the near-trivial case of
predicting that a user likes a certain book given that another book by the same author has been used for training.

\noindent{\bf Datasets:}
\label{subsec:datasets}
We use two book datasets, Goodreads  \cite{wan2018item} (\textbf{GR}) and Amazon books \cite{ni2019justifying} (\textbf{AM}), from the UCSD recommender systems repository \cite{McAuleyDatasets}.
Both datasets contain item titles, genre or category tags, item description, and user reviews.

Prior works mostly consider C-core data variants where all users and items have at least C interactions (C=10 or 5). This pre-processing focuses on interaction-based predictions, whereas our intention is to study the case of data-poor users and items. 
Instead, our data pre-processing is designed to evaluate text-based recsys performance with
text-rich users.
We select 1K users from each of the two datasets, based on descending order of average review length per book (filtered for English reviews).
We view all book-user interactions with a rating of 4 or higher as positive, and disregard the lower ratings as they are rare anyway.
We split the data into training, validation, and test sets (60:20:20), filtering out users with less than 3 items to guarantee at least one interaction per user in each set.
Table \ref{tab:dataset-stats} gives statistics for the datasets.
Both 1K-rich data variants are extremely sparse in terms of users that share the same items; so the emphasis is on leveraging text.


\noindent{\bf Baselines:}
We compare our approach to several state-of-the-art baselines, which cover different methods
, ranging from traditional collaborative filtering approaches to text-centric neural models {and LLM-based rankers}: 

\squishlist
    \item \textbf{CF}: collaborative filtering operating on user-item interaction matrix by 
    computing per-user and per-item vectors (dim=200) via matrix factorization~\cite{Funk-Netflix2006}.
    \item \textbf{LLMRank}: following \cite{hou2023large}, we use ChatGPT to rank the test items, given the user's reading history. The history is given by the sequence of titles of the 50 most recent books of the user, prefixed by the prompt ``I’ve read the following books in the past in order:''.
    This prompt is completed by a list of titles of test-time candidate items, asking the LLM to rank them.
    \item \textbf{P5-profile}:
    prompting the T5 language model \cite{raffel2020exploring}, 
    to provide a recommended item for a user, given their ids. Following \cite{DBLP:conf/recsys/Geng0FGZ22}, we train P5 using the prompts for \textit{direct recommendation} 
    to generate a \textit{``yes''} or \textit{``no''} answer. 
    Pilot experiments show that the original method does not work well on sparse data.
    Therefore, we extend P5 to leverage review texts and item descriptions. 
    Instead of ids, the prompts include item descriptions and sentences from reviews 
    with the highest idf scores (i.e., one of our own techniques).
    \item \textbf{BENEFICT}\cite{pugoy-kao-2020-bert}: uses a {frozen} BERT model to create representations for each user review, which are averaged and concatenated to the item vectors. Predictions are made by a feed-forward network on top.
    Following the original paper, each review is truncated to its first 256 tokens.
    \item \textbf{BENEFICT-profile}: our own variant of BENEFICT where the averaging over all user reviews is replaced by our idf-based selection of most informative sentences, with the total length limited to 128 tokens (for comparability CUP).
\squishend

\vspace{0.1cm}
\noindent{\bf Performance Metrics:}
Following the literature, we report NDCG@5 (Normalized
Discounted Cumulative Gain) with binary 0-or-1 gain and P@1 (precision at
rank 1). We compute these by micro-averaging over all test items of all users.
Macro-averaged results over users were not significantly different, hence they are not reported in the paper.
NDCG@5 reflects the observations that users care only about a short list of
top-N recommendations; P@1 is suitable for recommendations on mobile devices
(with limited UI). We also measured other metrics, like NDCG@k for higher k,
MRR and AUC. None of these provides any additional insight, so they are not
reported here.

\vspace{0.1cm}
\noindent{\bf Evaluation Modes:}
At test time, for each positive test item we sample 100 negative items from all unlabeled data. The system scores and ranks these 101 data points, creating a ranked list of items to be evaluated by the performance metrics introduced above.
We evaluate all methods in two different modes with respect to the negative sampling strategy:
\squishlist
\item {\bf Standard:} sampling the 100 negative test points uniformly at random.
\item {\bf Search-based:} given the positive test item, searching for the top-100 approximate matches to the item's description, 
using the  BM25 scoring model.
\squishend

\vspace{0.1cm}
\noindent{\bf Configurations:}
In the experiments, all CUP variants use an input budget of 128 tokens
as a stress test, emphasizing our goal of limiting the computational and energy costs.
The following CUP configurations cover different ways of user profile creation (see Subsection \ref{subsec:longtext}):

\squishlist
\item \textbf{CUP$_{idf}$}: review sentences selected by idf scores (weighted sentences). 
\item \textbf{CUP$_{sbert}$}: review sentences selected by similarity to the corresponding item description, using Sentence-BERT (similar sentences).
\item \textbf{CUP$_{1gram}$}: unigrams selected by
tf-idf scores {(weighted phrases)}.
\item \textbf{CUP$_{3gram}$}: 
3-grams selected by tf-idf scores {(weighted phrases)}.
\item \textbf{CUP$_{kwT5}$}: set of keywords generated by a fine-tuned 
T5 model.\footnote{\scriptsize{\url{https://huggingface.co/ml6team/keyphrase-generation-t5-small-inspec}}} 
\item \textbf{CUP$_{kwGPT}$}: 
a keyword profile generated by ChatGPT ({gpt-3.5-turbo}).
\item \textbf{CUP$_{kwLlama}$}: a keyword profile generated by 
Llama (\small{Meta-Llama-3.1-8B-Instruct}), given hand-crafted few-shot examples.
\item \textbf{CUP$_{absLlama}$}: an abstractive 1st-person narrative profile generated by Llama, given hand-crafted few-shot examples.

\squishend

For comparison, we also configure a simpler variant, {\bf CUP${tags}$}, which uses genre tags from the user's training items as the user text. 
To observe the effect of item metadata, we further restrict this variant to use only the item title and genre as item text, denoted as CUP${basic}$.

We used the following hyperparameters for CUP configuration, obtained though grid search: 4e-5 as learning rate, 256 as batch size, 200 as FFN size.
All methods were run on NVIDIA Quadro RTX 8000 GPU
with 48 GB memory, and
we implemented the models with PyTorch.
\vspace{-0.3cm}
\section{Experimental Results}
\vspace{-0.2cm}
\vspace{-1cm}
\begin{table}[h]
    \begin{minipage}{0.48\textwidth}
        \renewcommand{\arraystretch}{1}  

\caption{Standard evaluation.}
\label{finaltable:baselines-Standard}
\centering
\begin{adjustbox}{width=\textwidth}
\begin{tabular}{|l|ll|ll|} \hline
\multicolumn{5}{|c|}{\textbf{AM-1K-rich}} \\ \hline
&\multicolumn{2}{c||}{Train Uniform} & \multicolumn{2}{c|}{Train Weighted} \\ 
{Method} & NDCG@5 & {P@1} & NDCG@5 & {P@1} \\ \hline
{CF} & 3.06 & 1.0 & 2.88 & 0.69 \\ 
{LLMRank} & 4.62 & 2.27 & n/a & n/a \\ 
{BENEFICT} & 9.4 & 3.53 & 14.4 & 5.74 \\ 
\small{BENEFICT$_{prof}$} & 24.38 & 12.98 & 24.66 & 12.81 \\ 
{P5$_{prof}$} & 24.9 & 14.5 & n/a & n/a \\ 
{CUP$_{basic}$} & 25.42 & 13.64 & 26.95* & 14.27 \\ 
{CUP$_{tags}$} & 27.31* & 14.99* & 28.83* & 16.05* \\ 
{CUP$_{idf}$} & \textbf{29.21}* & \textbf{15.82*} & \textbf{31.09}* & \textbf{17.71}* \\ \hline 
\multicolumn{5}{|c|}{\textbf{GR-1K-rich}} \\ \hline
{CF} & 4.44 & 2.69 & 3.83 & 2.26 \\ 
{LLMRank} & 4.86 & 2.1 & n/a & n/a \\ 
{BENEFICT} & 23.76 & 12.17 & 25.23 & 13.26 \\ 
\small{BENEFICT$_{prof}$} & 30.26 & 16.83 & 31.77 & 17.74 \\ 
{P5$_{prof}$} & 28.01 & 14.46 & n/a & n/a \\ 
{CUP$_{basic}$} & 26.75 & 13.28 & 28.4 & 14.89 \\ 
{CUP$_{tags}$}& 30.92 & 16.3 & 33.28* & 17.83 \\ 
{CUP$_{idf}$} & \textbf{38.39}* & \textbf{22.26}* &\textbf{ 39.41*} & \textbf{22.01*}\\ \hline 
\end{tabular}

        \end{adjustbox}
        \renewcommand{\arraystretch}{1}
    \end{minipage}
    \hfill
    \begin{minipage}{0.48\textwidth}
        \renewcommand{\arraystretch}{0.95}  
\caption{Search-based evaluation.}
\label{finaltable:baselines-Search-based}
\centering
\begin{adjustbox}{width=\textwidth}
\begin{tabular}{|l|ll|ll|} \hline
\multicolumn{5}{|c|}{\textbf{AM-1K-rich}} \\ \hline
&\multicolumn{2}{c||}{Train Uniform} & \multicolumn{2}{c|}{Train Weighted} \\ 
{Method} & NDCG@5 & {P@1} & NDCG@5 & {P@1} \\ \hline
CF & 3.02 & 1.0 & 3.03 & 0.83\\ 
LLMRank & 3.49 & 1.1 & n/a & n/a \\ 
BENEFICT & 2.45 & 0.66 & 3.69 & 1.29 \\ 
\small{BENEFICT$_{prof}$} & 6.49 & 2.33 & 8.11 & 3.53 \\ 
P5$_{prof}$ & 8.4* & \textbf{3.88*} &  n/a &  n/a \\ 
CUP$_{basic}$ & 7.13 & 2.76 & 6.87 & 2.61 \\ 
CUP$_{tags}$ & 7.41 & 2.93 & 7.0 & 2.84 \\ 
CUP$_{idf}$ & \textbf{8.93*} & 3.53* & \textbf{9.14} & \textbf{4.02} \\ \hline
\multicolumn{5}{|c|}{\textbf{GR-1K-rich}} \\ \hline
CF & 3.73 & 2.02 & 3.25 & 1.74 \\ 
LLMRank & 4.57 & 1.79 & n/a & n/a \\ 
BENEFICT & 6.73 & 2.38 & 6.88 & 2.44 \\ 
\small{BENEFICT$_{prof}$} & 8.95 & 3.4 & 9.63 & 3.59 \\ 
P5$_{prof}$ & 9.15 & 3.01 & n/a & n/a \\ 
CUP$_{basic}$ & 9.35 & 3.46 & 9.84 & 3.55 \\ 
CUP$_{tags}$ & 10.66* & 4.3* & 11.18* & 3.94 \\ 
CUP$_{idf}$ & \textbf{14.18}* & \textbf{5.9}* & \textbf{13.76}* & \textbf{5.32}* \\ \hline
\end{tabular}
        \end{adjustbox}
        \renewcommand{\arraystretch}{1}
    \end{minipage}
\end{table}

\vspace{-1cm}

\subsection{Comparison of CUP against Baselines}
Table \ref{finaltable:baselines-Standard}
shows the results for the AM-1K-rich and GR-1K-rich data,
comparing our default configuration CUP$_{idf}$
against all baselines, for the two different ways of sampling negative training points. 
Results with statistical significance over the 
BENEFICT$_{prof}$ baseline,
by a paired t-test with p-value~$<$~0.05, are marked with an asterisk.
Bonferroni correction for multi-hypotheses testing is applied, reducing the 
test level
of each pairwise comparison to 0.005.
We make the following key observations:

\squishlist
\item The interaction-centric CF 
fails for this extremely sparse data. 
BENEFICT, utilizing the entire user review texts, 
shows poor performance.
At the same time, BENEFICT$_{prof}$ and  P5$_{prof}$, 
extended with our text-derived profiling, achieve decent performance.

\item LLMRank 
also performs poorly.
Solely relying on the LLM's latent knowledge about books is not sufficient
when coping with long-tail items.
Popularity and position bias \cite{hou2023large} further aggravate this adverse effect.
To mitigate position bias,
we ran a variant with smaller test sets of only 20 candidate {negative} items (per positive test item), as in the original setup of \cite{hou2023large}.
This boosts the NDCG@5 for LLMRank from 4.8\% to 23.5\%, on GR-1k-rich in standard evaluation, which is still a large margin below CUP$_{idf}$ reaching 38.3\% (and similarly big gaps for the other dataset).  
\item Between the three CUP configurations, we see a clear trend: 
review-based user profiling ($idf$) outperforms user profiles based on genres or categories ($tags$), and simplifying item-side text to titles and tags alone by removing item description ($basic$) performs worst.
Remarkably, even CUP$_{basic}$ is better than all the baselines. CUP$_{idf}$ is ca. 5 percentage points better in NDCG@5 than the baselines.
\item In search-based evaluation 
(Table \ref{finaltable:baselines-Search-based}), 
the absolute results are much lower, emphasizing the difficulty of this realistic mode. Still, the relative comparisons between methods are nearly identical to the results with standard evaluation. 
Again, CUP$_{idf}$ is the winner, with a clear margin.
\item The absolute numbers on GR-1K-rich are generally higher, due to the different
data characteristics. The gains by CUP$_{idf}$ over the baselines and
over the simpler CUP configurations are even more pronounced
(e.g., outperforming BENEFICT$_{prof}$ by 8 percentage points with
standard evaluation).
\squishend

\vspace{-0.4cm}
\vspace{-0.1cm}
\subsection{Efficiency of CUP}
\label{subsec:experiments-efficiency}
\vspace{-0.2cm}
Two architectural choices make CUP efficient:
\squishlist
    \item input length restricted to 128 tokens, and
    \item fine-tuning only the last layer of BERT and the FFN layers.
\squishend

\begin{figure}[h]
    \centering
    \begin{minipage}{0.54\textwidth}
        \centering
        \captionof{table}{CUP results, by user/item groups (NDCG@5 with Search-based evaluation).}
        \label{finaltable:CUP-Search-based}
\resizebox{\textwidth}{!}{
\begin{tabular}{|l|lllllll|} \hline
\multicolumn{8}{|c|}{\textbf{AM-1K-rich}} \\ \hline
Method & {ALL} & {u-s} & {u-r} & {u-b} & {s-s} & {s-r} & {s-b} \\ \hline 
CUP$_{idf}$ & 9.14 & 5.93 & 7.09 & 12.0 & 8.6 & 11.71 & 12.78 \\ 
CUP$_{sbert}$ & 9.0 & 5.19 & \textbf{7.32} & 11.46 & 9.44 & 14.53 & 14.23 \\ 
CUP$_{1gram}$ & 9.08 & \textbf{6.24} & 7.14 & 11.21 & \textbf{9.76} & \textbf{17.0} & 13.25 \\ 
CUP$_{3gram}$ & 8.98 & 5.54 & 7.01 & 11.81 & 8.2 & 13.97 & 10.99 \\ 
CUP$_{kwT5}$ & \textbf{9.5} & 6.03 & 7.15 & \textbf{12.48} & 8.71 & 12.38 & \textbf{17.78} \\ 
CUP$_{kwGPT}$ & 8.93 & 5.95 & 6.95 & 11.59 & 8.29 & 11.3 & 13.85 \\ 
CUP$_{kwLlama}$ & 9.01 & 6.23 & 6.9 & 11.82 & 7.83 & 11.15 & 12.45 \\ 
CUP$_{absLlama}$  & 8.04 & 5.14 & 5.86 & 10.63 & 5.23 & 16.11 & 11.94 \\ \hline  
\multicolumn{8}{|c|}{\textbf{GR-1K-rich}} \\ \hline

CUP$_{idf}$ & 13.76 & 7.32 & \textbf{11.42} & 14.96 & 17.37 & 17.43 & 19.56 \\ 
CUP$_{sbert}$ & 13.51 & \textbf{8.47} & 10.41 & 15.05 & 16.14 & 16.85 & 19.79 \\ 
CUP$_{1gram}$ & 13.47 & 7.82 & 10.61 & 14.64 & 16.14 & 17.95 & 20.32 \\ 
CUP$_{3gram}$ & 13.19 & 7.3 & 10.47 & 15.08 & 14.06 & 16.16 & 17.94 \\ 
CUP$_{kwT5}$ & 13.76 & 7.51 & 11.03 & 14.43 & 17.43 & 19.33 & 22.17 \\ 
CUP$_{kwGPT}$ & \textbf{13.78} & 7.97 & 10.56 & 14.33 & 18.32 & \textbf{20.36} & \textbf{23.18} \\ 
CUP$_{kwLlama}$ & 13.54 & 8.06 & 10.63 & \textbf{15.25} & 13.88 & 16.83 & 19.54 \\ 
CUP$_{absLlama}$ & 13.22 & 6.75 & 9.87 & 14.2 & \textbf{18.76} & 19.59 & 21.37 \\ \hline

\end{tabular}
}
    \end{minipage}%
    \hfill
    \begin{minipage}{0.45\textwidth}
        \centering
     \includegraphics[width=\linewidth]{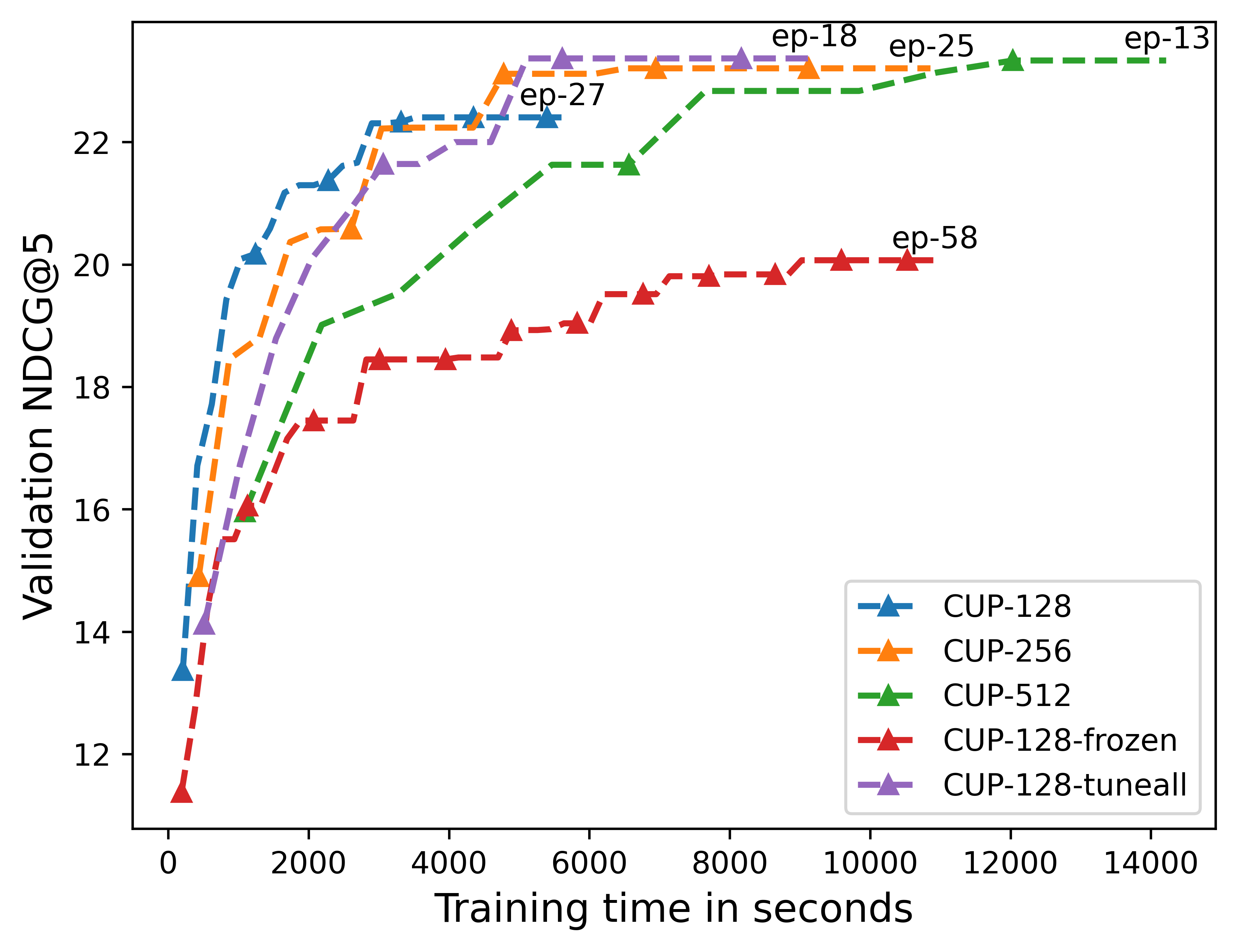}  
\caption{Training time for different input lengths and trainable parameters (lines are marked every 5th epoch).}
  \label{fig:timecomparisons}
    \end{minipage}
\end{figure}

For more analysis, we measured the training time and resulting NDCG on the
{GR-1K-rich} data, comparing different input size budgets and 
{choices of tunable parameters.}
Figure~\ref{fig:timecomparisons} shows the {NDCG@5} results {on the validation set.}

We observe that the 128-token configuration has the lowest training cost: significantly less time per epoch than the other variants and fast convergence (reaching its best NDCG already after 15 epochs in ca. 3000 seconds). The 256- and 512-token models eventually reach higher NDCG, but only by a small margin and after much longer training time.

As for the number of trainable parameters, 
we observe that the variant with frozen BERT takes much longer to converge and is inferior to the preferred CUP method even after more than 50 epochs.
The other extreme, allowing all BERT parameters to be altered,
performs best after enough training epochs. 
However, it takes almost twice as much time per epoch. From the benefit/cost perspective, our design choice hits a sweet spot in the spectrum of model configurations.

\vspace{-0.3cm}
\vspace{-0.1cm}

\vspace{-1cm}

\subsection{Comparison of CUP Configurations}
For insight on 
specific groups of users and items, we split the 1000 users and their items into the following groups,
reporting NDCG@5 
for each
group separately.
Note that this refinement drills down on the test outputs; the training is unaffected. 
\squishlist
\item Items are split into {\bf unseen (u)} and {\bf seen (s)} items.
Unseen test items have not been seen at training time. Seen test items appeared as positive training items for a different user.
\item Users are split into groups based on 
the {\large $\sharp$}books-per-user distribution:
\squishlist
\item {\bf Sporadic (s)} users 
are the lowest 50\% with the least numbers of books. 
For GR-1K-rich, this threshold is 13 books per user; for AM-1K-rich it is 5 (with means 6 and 3, resp.). 
\item {\bf Regular (r)} users are those between the 50th percentile and 90th percentile, which is between 13 and 71 books per user for 
GR-1K-rich, and between 5 and 20 for AM-1K-rich
(with means 31 and 9, resp.).
\item {\bf Bibliophilic (b)} users 
are the highest 10\%: above 75 {books per user} for GR-1K-rich and above 20 for AM-1K-rich (with means 156 and 43, resp.).
\squishend
\squishend

Table \ref{finaltable:CUP-Search-based}  shows NDCG@5 (for all and per item/user group) with search-based mode, comparing all CUP configurations 
with weighted training.
 We offer the following notable observations:
 \squishlist
\item 
Across all groups
, all CUP configurations are competitive. The overall differences between them are relatively small. The winner, by a small margin, is CUP$_{kwT5}$ for AM and CUP$_{kwGPT}$ for GR datasets, closely followed by the default configuration CUP$_{idf}$ and CUP$_{kwLlama}$ as well as  CUP$_{sbert}$. None of the methods is able to extract the ``perfect'' gist from the noisy review texts; but all of them do a decent job. Despite the fact that generated profiles are slightly ahead of the others, the bottom line is that a relatively simple configuration, like idf-selected sentences, is a good choice.

 \item The CUP$_{kwGPT}$ variant achieves its highest gains for the richer item/user groups: seen items and regular or bibliophilic users (GR dataset). This provides ChatGPT with longer and more informative texts. A similar effect, but to a lesser and noisier extent, can be observed for T5-based CUP$_{kwT5}$. Conversely, these methods perform substantially worse on the sporadic-unseen group. 
 \squishend

\vspace{-0.2cm}
\renewcommand{\arraystretch}{1.1}
\begin{table}[]
    \centering
    {\footnotesize
    \begin{tabular}{|l|ccc|cc|ccc|cc|} \hline
         & \multicolumn{5}{c|}{Amazon} & \multicolumn{5}{c|}{Goodreads} \\ \hline
         & \multicolumn{3}{c|}{POS} &  &  & \multicolumn{3}{c|}{POS} &  & \\
         & NN & VB & AD & avg. idf & KL-div & NN & VB & AD & avg. idf & KL-div \\ \hline
all reviews         & 0.25 & 0.19 & 0.17 & 0.009 & 3.28        & 0.28 & 0.19 & 0.16 & 0.008 & 2.93    \\ 
idf sentences       & 0.33 & 0.16 & 0.18 & 0.008 & 1.72        & 0.39 & 0.15 & 0.19 & 0.011 & 1.19    \\ 
Llama abstractive   & 0.35 & 0.15 & 0.18 & 0.065 & 1.7         & 0.36 & 0.14 & 0.19 & 0.084 & 1.31    \\ 
Llama keywords      & 0.67 & 0.05 & 0.23 & 0.42 & 1.55        & 0.68 & 0.04 & 0.23 & 0.503 & 1.27     \\ 
unigrams            & 0.47 & 0.23 & 0.25 & 0.537 & 1.99        & 0.54 & 0.18 & 0.22 & 0.921 & 1.22    \\ \hline
    \end{tabular}
    }
    \caption{Statistical comparison of different kinds of user profiles}
    \label{tab:profile_stats}
\end{table}
\renewcommand{\arraystretch}{1}
\vspace{-1.5cm}

\section{Analysis of User Profiles}
\label{sec:profile-analysis}

Desiderata for an ideal profile are
i) {\em high utility}: leading to strong recommender performance,
ii) {\em easy interpretability:} supporting humans in understanding the gist of somebody's interests, and
iii) {\em sound faithfulness:} capturing the user's style in writing reviews.
Clearly, there are trade-offs between these dimensions.
In terms of utility, our experiments show that several kinds of profiles are roughly on par, with some simple ones being slightly ahead. On the other hand, the most interpretable profiles are the generative ones using an LM.
Finally, the extractive profiles, like salient sentences from reviews, 
appear most faithful.
For illustration, Table \ref{table:profiles} provides examples of different kinds of profiles.

To obtain more insights,
we computed various statistics: distributions of part-of-speech (POS) tags (i.e., word categories) and distributions of idf-weight mass among frequent words.
These are derived by
concatenating profiles of all
users, for each of the most interesting profile types, including the users' original reviews.
 
The statistics are shown in Table \ref{tab:profile_stats}.
The first column denotes the fractions of the three most frequent POS types. We observe that the LM-generated profiles have a much higher share of nouns, which are more informative than verbs or adjectives/adverbs. 
The second column shows 
the {average} idf weight for the words in the 20-th percentile 
of the word frequency distribution.
We observe that the original reviews carry low idf weight, reflecting their verbose and noisy nature. In constrast, the generative profiles 
select high-idf words as most informative cues.
Finally, the third column shows the 
Kullback-Leibler divergence (with Laplace smoothing, $\alpha=0.1$)
for the user profiles generating the book descriptions (with removal of stopwords).
We observe that the vocabulary of original reviews differs significantly from the wording in book descriptions, whereas idf-aware profiles and keyword-generated profiles are much closer to the vocabulary that describes book contents.


\vspace{-0.2cm}
\section{Conclusion}
\vspace{-0.2cm}
This work
presented a transformer-based framework CUP, with novel techniques for constructing concise user profiles by judiciously
selecting informative pieces of user text.
Our experiments, with both standard evaluation and a search-based mode, show that leveraging user text is beneficial in this data-poor regime,
and that CUP methods clearly outperform
state-of-the-art baselines like BENEFICT, P5, and LLM-Rec. 
Among the CUP configurations with different profiles, we observed that most perform similarly. This suggests two main options in practice: choose the lowest-cost variant which uses idf-based n-grams or sentence-level excerpts of reviews, or choose the ones that are generated by an LLM, at higher cost, if use cases prioritize the human-readability and ability to edit profiles.

\begin{table}[tbh]
\centering
\caption{User profiles constructed by various methods (truncated to max 3 lines).}
\label{table:profiles}
{\footnotesize
\renewcommand{\baselinestretch}{0.7} 
\begin{tabular}{lm{11cm}} 
    \textbf{Method} & \multicolumn{1}{c}{\textbf{User Profile}} \\ \hline \hline
    \textbf{genres} & \small africa, nature ecology, americas, history, europe, leaders notable people, relationships, world, science math, historical, biographies memoirs, self - help, genre fiction, literature fiction
    \\ 
    \hline 
    \parbox{2cm}{\textbf{idf} \\ \textbf{sentences}}  & \small Confederate Navy Raider. Irish history - in a nutsell!. Heroes, US Marine Corps Medal of Honor Winners by Marc Cerasini. Women's Options on the American Fronteir. Norwegian Immigration 1850s. Life on the Praire. Nice Quirky Book.
    \\     \hline 
    \parbox{2cm}{\textbf{SBERT} \\ \textbf{sentences}}&\small It was actually mostly written by Elisabeth Koren, wife of Reverend U. V. Koren. The area of research is the Arkansas Missouri Borderlands. Story is of a woman radical whose life was brought up short by Senator McCarthy during
    \\     \hline 
    \textbf{unigrams}&\small texas, navy, colt, marines, koren, norwegian, quege, ranger, vilhelm, norwegians, nutsell, leponto, markist, rangers, caliber, fronteir, cerasini, korens, book, cavalry, laundress, pistols, xo, agnes, praire, outstanding, army
    \\ \hline
    \parbox{2cm}{\textbf{T5} \\ \textbf{keywords}}  & \small  norweger immigration 1850s book, texas, texas, funeral. amateur historian, history buffs, acoustic borderlands, ignoble deeds, south. socialist woman, gender issues, birth control, abortion, labor unions, health care, social care
    \\ \hline
    \parbox{2cm}{\textbf{ChatGPT} \\ \textbf{keywords}} & \small norwegian immigration, 1800s life, historical commentary, civil war, confederate navy, mexican war, texas cavalry, american frontier, lost states, irish history, texas rangers, mexican war, agnes smedley, 
    norwegian female lives
    \\ \hline 
    \parbox{2cm}{\textbf{Llama} \\ \textbf{keywords}} & \small Norwegian Immigration, 1850s, historical reference, Civil War, eyewitness accounts, Arkansas Missouri Borderlands, Confederate Navy Raider, Mexican War, Fort Brown, Western Frontier, Simon Kenton, D. Boone, Crusades
    \\ \hline
    \parbox{2cm}{\textbf{Llama} \\ \textbf{1st-person}} & \small I'm a history buff with a passion for non - fiction books, particularly those that delve into the lives of ordinary people during extraordinary times. I enjoy reading about the experiences of women, immigrants, and marginalized groups
    \\ \hline
\end{tabular}
}
\end{table}
\vspace{-2cm}

%
%
\clearpage
\bibliographystyle{splncs04}
\bibliography{ref}

\end{document}